\documentclass{cpbtex}
\usepackage[version=4]{mhchem}

\usepackage{geometry}%页面设置
\usepackage{graphicx}%图片设置

\usepackage{threeparttable}
\usepackage{float}
\usepackage{subfig}%多个子图
\usepackage{caption}%注释设置
\usepackage[normalem]{ulem}
\usepackage{color}
\usepackage{color,framed}%可以中文
\usepackage{hyperref}
\usepackage[hyphenbreaks]{breakurl}

\begin{document}
%\begin{CJK*}{UTF8}{song}

\newcommand{\gem}[0]{\tilde{\mathcal R}}
\newcommand{\ebd}[0]{\tilde{\mathcal G}}

\newcommand{\Al}[0]{\textrm{Al}}
\newcommand{\Mg}[0]{\textrm{Mg}}
\newcommand{\Cu}[0]{\textrm{Cu}}
\newcommand{\formation}[0]{\textrm{af}}
\newcommand{\recheck}[1]{{\color{red} #1}}

\title{Accurate Deep Potential model for the Al-Cu-Mg alloy in the full concentration space \thanks{The work of H. W.
is supported by the National Science Foundation of China under Grant No. 11871110, the National Key Research and Development Program of China under Grants No. 2016YFB0201200 and No. 2016YFB0201203, and Beijing Academy of Artificial Intelligence (BAAI). The work of L. Z. was supported in part by  the Center of Chemistry in Solution and at Interfaces (CSI) funded by the DOE Award de-sc0019394.}}

\author{Wanrun Jiang$^{1,2}$, \ Yuzhi Zhang$^{3,4}$, \\ Linfeng Zhang$^{5,}$\thanks{Corresponding author. E-mail: linfengz@princeton.edu}~, \ and \ Han Wang$^{6,}$\thanks{Corresponding author. E-mail:~wang\_han@iapcm.ac.cn }\\
$^{1}$Songshan Lake Materials Laboratory,\\Dongguan, Guangdong 523808, People's Republic of China\\ $^{2}$Institute of Physics, Chinese Academy of Sciences,\\ Beijing 100190, People's Republic of China\\  % The line break was forced via \\
$^{3}$Beijing Institute of Big Data Research,\\ Beijing 100871, People's Republic of China\\
$^{4}$Yuanpei College of Peking University,\\
Beijing 100871, People's Republic of China
\\ % The line break was forced via \\
$^{5}$Program in Applied and Computational Mathematics,\\
Princeton University, Princeton, NJ 08544, USA\\
$^{6}$Laboratory of Computational Physics,\\
Institute of Applied Physics and Computational Mathematics,\\
Huayuan Road 6, Beijing 100088, People's Republic of China\\
}  
\maketitle
\newpage
\begin{abstract}
Combining first-principles accuracy and empirical-potential efficiency for the description of the potential energy surface (PES) is the philosopher's stone for unraveling the nature of matter via atomistic simulation.
This has been particularly challenging for multi-component alloy systems due to the complex and non-linear nature of the associated PES.
In this work, we develop an accurate PES model for the Al-Cu-Mg system by employing Deep Potential (DP), a neural network based representation of the PES, and DP Generator (DP-GEN), a concurrent-learning scheme that generates a compact set of {\it ab initio} data for training.
The resulting DP model gives predictions consistent with first-principles calculations for various binary and ternary systems on their fundamental energetic and mechanical properties, including formation energy, equilibrium volume, equation of state, interstitial energy, vacancy and surface formation energy, as well as elastic moduli.        
Extensive benchmark shows that the DP model is ready and will be useful for atomistic modeling of the Al-Cu-Mg system within the full range of concentration.

\end{abstract}


%Computational modeling; simulation From section Interdisciplinary_MS\\

\section{Introduction}
Aluminum, copper and magnesium (Al-Cu-Mg) based alloys are among the most versatile metallic materials that meet all the requirements of lightweight, high strength and good fatigue resistance.
They are intensively used in automotive, aviation, aerospace and engineering machinery industries, and expend the scope of activity and the production efficiency of human civilization.\textsuperscript{\cite{dursun2014recent}}
The industrial application of Al-Cu-Mg alloys could be traced back to 1906, when age hardening, as the key phenomenon enhancing their specific strength, was first discovered in this material.\ucite{Wilm1911Alalloy}
It led to the well known wrought alloy, duralumin, and soon enabled structural members of Zeppelin airships and the world first passenger airplane without frame or bracing wire to support its wings.
After a century, derived commercial aluminum alloy 2000 series are still one of the major material families used for aircraft manufacture.\textsuperscript{\cite{WarrenBoeing}}
Decades of explorations have linked up the conceptual road-map of age hardening in Al-Cu-Mg alloys, giving a mainline from the super saturated solid solution to a sequence of precipitation that finally impede the movement of dislocations.\textsuperscript{\cite{wang2005precipitates,SigmundPrecipitatesX,EsinTwoOnePrecipitation}}
The presence, morphology and competition of micro-structures were observed by experiments,\textsuperscript{\cite{RingerZeroEightPrecipitation,marceau2010solute,STYLES201564,EsinTwoOnePrecipitation}} while their chemical formula, atomic configurations and relative stability were informed by theoretical calculations.\textsuperscript{\cite{liu2011the,zhang2012structural,SigmundPrecipitatesX}}
Phenomenological experiences have been accumulated along the road-map, revealing further complexity of the phenomenon
including, but being not limited to, the involvement of distinguishing processes like dislocation-solute interaction and solute clustering in different ageing stages,\textsuperscript{\cite{marceau2010evolution,NAGAIZeroOnevac}} the evolution of micro-structures under variant ageing conditions,\textsuperscript{\cite{PARELOneZeroSTwo,lin2012precipitation}} the ageing improvements on desired performance like dimensional stability\textsuperscript{\cite{SONGOneSeventwostage}} and variants of related processes in more complex scenario like Al-Cu-Mg nanocrystals.\textsuperscript{\cite{CHENOneSixNanoCry}}

A better understanding of the fundamental principles underlying the experiments needs the atom-resolved tracking of the dynamical processes, which is usually beyond the capability of current experiment equipment.
Therefore, molecular simulation techniques like molecular dynamics (MD) is a promising alternative, that is in principle capable to produce atom-scale information of the physical quantities of interest in the investigated processes.
MD simulations have provided insights into the precipitation strengthening of Al-Cu alloys, which mechanisms might also be inspiring for understanding related processes in Al-Cu-Mg systems. 
For example, the dislocations-precipitate interactions, that determine the strengthening effect, are dominated by looping process for most edge dislocation cases\textsuperscript{\cite{SinghOneZeroMechanisms}} and are significantly influenced by the temperature-dependant cross-slip for screw dislocations.\textsuperscript{\cite{SINGHOneOneCrossSlip}}
The transition among meta-stable precipitates could be adjusted by the vacancy segregation, while the vacancy densities could further be controlled by heating the surface of the designed nanoscale specimen or exposing bulk counterparts to intense irradiation.\textsuperscript{\cite{BourgeoisTwoZerotransforming}}
Besides, possible common factors for precipitation strengthening in different alloys could also be perceived by MD simulations, such as the importance of local interface curvature in driving the dislocation-precipitate interactions.\textsuperscript{\cite{PRAKASHOneFiveCurvature}}

Ideally, \textit{ab initio} molecular dynamics (AIMD),\textsuperscript{\cite{car1985unified}} which evolves atomic systems under the Newton's law of motion with forces generated on-the-fly by first-principles-based methods,
holds the promise to provide an accurate description of Al-Cu-Mg alloy systems.
However, due to the high computational cost, first-principles-based methods like density functional theory (DFT)\textsuperscript{\cite{kohn1965self}} can only be used to simulate materials within the spacial scale of a few hundreds of atoms and the temporal scale of a few dozens of picoseconds, using desktop workstations or ordinary computer clusters.
Despite successes in computing basic quantities like structural factors, heat of formation, and cohesive energy for important precipitation,\textsuperscript{\cite{liu2011the,BOURGEOIS20117043,zhang2012structural}}
they are in general not computationally feasible to reveal the atomistic mechanism in large-scale dynamic processes like phase transition, precipitation, and interface migration under working conditions, whose spacial-temporal scales are far beyond the current capability of AIMD.

In these cases, one has to resort to empirical force fields (EFFs).
Representative examples include the Lennard-Jones potential,\textsuperscript{\cite{jones1924LJ}} the Stillinger-Weber potential,\textsuperscript{\cite{stillinger1985SW}} the embedded-atom method (EAM) potential\textsuperscript{\cite{daw1984EAM}} and the modified embedded atom method (MEAM) potential.\textsuperscript{\cite{baskes1992modified}}
By approximating the potential energy surface (PES) with relatively simple and analytical functions and optimizing the parameters with a group of target properties, EFF-based MD is orders of magnitude more efficient than AIMD.
However, it has been challenging to construct EFF models for multi-component systems like the  Al-Cu-Mg alloy. 
The difficulties are two-folds: (1) The relatively simple function form makes it difficult to fit all target properties in the full concentration space, especially in the case of ternary alloy. 
(2) The number of tunable parameters in the EFF grows quadratically with respect to the number of components of the system.
The number of parameters in an $M$-component system is $M(M+1)/2$ times larger than that of a single-component system, so the optimization of the parameters becomes more difficult, if not impossible, for multi-component systems. 

For the Al-Cu-Mg system, most of the existing EFF models are restricted to pure metals\textsuperscript{\cite{pascuet2015atomic,choudhary2015charge,mendelev2013the,asadi2015two-phase,etesami2018molecular,zhou2004misfit-energy-increasing,sun2006crystal-melt,wilson2016a}} and binary components.\textsuperscript{\cite{liu1999a,apostol2011interatomic,zhou2016an,LIU1997357,liu1998grain-boundary,mendelev2009development}}
Although some of them have been demonstrated to provide decent descriptions of the Al end in the Al-Cu phase diagram\textsuperscript{\cite{zhou2016an}} and the solid–liquid phase equilibria in the target component range of Al-Mg alloy,\textsuperscript{\cite{mendelev2009development}} the generalization of these single or binary component EFFs to ternary Al-Cu-Mg alloy is not straightforward.
One of the few examples of multi-component EFF\textsuperscript{\cite{jelinek2012modified}} that covers Al, Cu, and Mg is constructed within the formalism of MEAM potential,\textsuperscript{\cite{baskes1992modified}} and is explicitly optimized for single-element systems of Al, Si, Mg, Cu, and Fe, their binary pairs in a hypothetical NaCl reference structure, as well as some stable binary intermetallic components.
The fitting targets include a group of selected properties obtained from \textit{ab initio} calculations, such as generalized stacking fault energy for single elements, heat of formation, equilibrium volume, elastic moduli of hypothetical binary pairs, and heat of formation for some stable binary components.
The transferability of this MEAM potential to the full concentration space is implicitly suggested, but was found not satisfactory for fundamental properties partially beyond the fitted sets for developing the potential.\textsuperscript{\cite{zhang2019active}}
Therefore, there is a demand for a PES model that is substantially less expensive than \textit{ab initio} methods, and, in the meanwhile, provides remarkably higher accuracy and wider applicable range than the state-of-the-art EFFs.
Ideally, it should describe the static, mechanical, and dynamical properties with a satisfactory accuracy in the full concentration space of the ternary Al-Cu-Mg system.

Recent advances in machine learning (ML) techniques\textsuperscript{\cite{behler2007generalized, bartok2010gaussian,schutt2017schnet,han2017deep,zhang2018deep,zhang2018end}} provide a promising route towards an accurate and efficient PES model for ternary Al-Cu-Mg alloys.
Here we consider the smooth edition of Deep Potential (DP),\textsuperscript{\cite{zhang2018end}} a deep learning based PES representation. 
It has been demonstrated that, when trained with \textit{ab initio} data, DP agrees well with an \textit{ab initio} PES and has an efficiency comparable to that of EFFs. 
The representation ability of DP stems from the outstanding ability of the deep neural networks to approximate high-dimensional functions,\textsuperscript{\cite{barron1993universal,ma2019machine}} and from a group of automatically generated symmetry-preserving descriptors, which faithfully capture the information lying in the local environment of atoms.
Compared with most ML models, the flexible and trainable descriptors of DP make it more suitable for complex tasks, e.g., multi-component systems like \ce{TiO2}-\ce{H2O} interface,\textsuperscript{\cite{andrade2020free}} strengthening precipitates from aluminum alloys,\textsuperscript{\cite{BourgeoisTwoZerotransforming}} as well as a \ce{(Zr_{0.2}Hf_{0.2}Ti_{0.2}Nb_{0.2}Ta_{0.2})C} high entropy alloy system with six chemical species.\textsuperscript{\cite{dai2020theoretical}}
To construct the DP model with an optimal set of data, a concurrent learning strategy, the Deep Potential Generator (DP-GEN),\textsuperscript{\cite{zhang2019active}} is further established.
Implemented as a close-loop iterative workflow, DP-GEN could generate the most compact and adequate data set that guarantees the uniform accuracy of DP in the explored configuration space. 
The effectiveness and efficiency of DP-GEN have been validated in the cases of pure Al, pure Mg, and the Al-Mg alloy\ucite{zhang2019active}, as well as pure Cu,\textsuperscript{\cite{zhang2020dpgen}} where the DP models outperform the state-of-the-art MEAM potentials in almost all examined properties.

In this paper, we construct the DP model for the Al-Cu-Mg alloys using the DP-GEN scheme for the whole concentration space, i.e.~\ce{Al_x Cu_y Mg_z} with $0\leq x, y, z \leq 1, x+y+z = 1$,
and the configuration space covering a temperature range around 
50.0 to 2579.8~K and a pressure range around 1 to 50000~bar. 
The resulting model gives generally better consistency with DFT results than the state-of-the-art MEAM model\textsuperscript{\cite{jelinek2012modified}} when predicting basic energetic and mechanical properties for systems distributed in the full concentration space of Al-Cu-Mg alloy.
The tested systems are taken from the Materials Project (MP) database, and most of them are not explicitly covered by the training set.
The tested properties include formation energy and volume for equilibrium state, equation of state (EOS), elastic modulus, unrelaxed surface formation energy for both binary and ternary alloys, as well as the relaxed vacancy formation energy and interstitial energy for ternary alloys.

In the following, we will first give a brief introduction of the DP model and DP-GEN scheme for the Al-Cu-Mg system in Sec.~\ref{sec:method}~~~
Then in Sec.~\ref{sec:results}~~~, we will focus on the comparison of various properties predicted by DFT, DP, and MEAM. 
Finally, a conclusion will be given in the last section.

\section{Method}
\label{sec:method}

The DP-GEN scheme works in an iterative manner.
Each iteration is composed of three steps, exploration, labeling, and training. 
With an initial guess of an ensemble of DP models generated from a relatively simple initial data set, DP-GEN starts to explore the configurational and chemical space by a sampler driven by DP-based simulation.
The explored configurations are then classified as accurate, candidate, and failed, according to the \emph{model deviation} defined as the standard deviation of the atomic force predictions of an ensemble of models. 
Models in the ensemble use the same neural network architecture and are trained on the same data set.
The only difference among them is the random seeds used to initialize parameters of the neural network.
Thus, their predictions would converge if the configuration has been well covered by current data set and deviate from each other otherwise.
The accurate set consists of the configurations with the \emph{model deviation} lower than a user-provide lower bound. The failed configurations are those with the \emph{model deviation} higher than a user-provided upper bound.
The failed configurations are typically generated by the DP models with a relatively poor quality in early DP-GEN iterations, and are likely to be unphysical, e.g.~with overlapping atoms.
A subset of the candidate configurations with errors between the lower and upper bounds are then sent to the labeling step, at which the labels of a configuration, i.e.~the energy, the virial tensor, and the atomic forces, are computed by DFT. 
Finally, the labelled configurations are added to the existing data set.
From the updated data set a new ensemble of DP models are trained, and a new iteration starts from the exploration step. 
The DP-GEN iterations converge when almost all the explored configurations are classified as accurate.
More details of the DP-GEN scheme are referred to previous works.\textsuperscript{\cite{zhang2019active,zhang2020dpgen}}

It should be remarked that
(1) The quality of the final DP model is insensitive to the data set that constructs the initial DP model to start the DP-GEN scheme. 
However, with a poor initial guess of DP, more labelled configurations from later iterations would be needed to improve the model. 
(2) The conditions under which the DP model is  used should be determined by the user beforehand, and the exploration strategy is accordingly designed to ergodically explore the relevant configuration space.
In this context, the sampler can be, but is not restricted to, molecular dynamics, Monte-Carlo simulation, structure prediction schemes, or enhanced sampling methods. 
(3) By classifying and sub-sampling the explored configurations, only the configurations of low accuracy are sent to the labeling step, so the most compact data set, i.e.~the minimal data set that helps DP achieve a certain accuracy, can be generated during the DP-GEN iterations.

The DP-GEN iterations are conducted by the open-source package DP-GEN.\textsuperscript{\cite{zhang2020dpgen}}
More details on initial data set, exploration, labeling, and training steps are provided as follows.

\paragraph{Initial data set.}
The initial data set used to kick-off the training and exploration can be generated through an automatic workflow integrated in the DP-GEN scheme, which allows the whole scheme to start from little prior knowledge. 
For alloys, four procedures are sequentially conducted. First, for each user-specific crystalline structure and each possible concentration,
the atomic species is randomly assigned on the lattice points.
Then, random perturbations are performed on the coordinates by adding values drawn from a
uniform distribution in the range $[-p_a,p_a]$. 
Perturbations are also performed on the cell vectors by a symmetric deformation matrix that is constructed by adding random noise drawn from a uniform distribution in the range $[-p_c, p_c]$ to an identity matrix.
Next, starting from the perturbed structures, AIMD simulations are conducted to produce labelled data with DFT calculated energy, force, and virial tensor.
Finally, the labelled data are used to create the initial data sets.
The preparation for single-element systems follows the same procedure without the need to consider possible concentrations.
Further details for preparing initial data sets can be found in previous reports.\textsuperscript{\cite{zhang2019active,zhang2020dpgen}} 
It is worth noting that, users can generate and collect initial data sets through other appropriate manners and tools. 
The procedure mentioned above just provides a convenient protocol and the quality of the finial model is not sensitive to the initial data sets.

In this work, for binary and ternary alloys, we consider body-centered-cube (BCC), hexagonal-close-packed (HCP), and face-centered-cube (FCC) structures, within $2\times 2\times 2$ super-cells.
For perturbations, $p_a$ is set to 0.01~{\AA} and $p_c$ is set to 3\%.
For single-element cases, training data sets for copper are inherited from previous work,\textsuperscript{\cite{zhang2020dpgen}} while those for aluminium and magnesium are prepared through individual single-element DP-GEN procedure in similar strategies. 
Besides, scaling factors compressing the cell are used in above procedure for binary and ternary alloys. 
Generated labelled data are added to the final data sets to further cover the high-pressure regime of the configuration space. 
Specific values are 0.84 to 0.98 with the interval of 0.02 for binary alloys and 0.85 to 0.97 with the interval of 0.03 for ternary alloys.

\paragraph{Exploration.}
In this work, the configuration and chemical space are explored by DP-based molecular dynamics (DeePMD) using the LAMMPS package.\textsuperscript{\cite{plimpton1995lammps}}
Two types of structures, bulk and surface, are considered. 
For bulk systems, the initial configurations are constructed from randomly perturbed FCC, BCC, and HCP, in $2\times 2\times 2$ super-cells with lattice positions randomly occupied by Al, Mg, or Cu atoms. 
It is noted that the $2\times 2\times 2$ super-cells are large enough to capture the local atomic environments needed to accurately construct the DP model for the Al-Cu-Mg alloys (see the Supplementary Text, Fig.~\ref{fig:Potentia energy 333} and Fig.~\ref{fig:accu ratio} for the accuracy of the model when predicting alloy structures with larger super-cells up to $12\times 12\times 12$).
Isothermo-isobaric ($NpT$) MD simulations are conducted at thermodynamic states ranging from 50.0 to 2579.8~K (twice the experimental melting point of Cu) and from 1 to 50000~bar. 
In the first 40 iterations (0-39), binary alloys (Al-Cu, Al-Mg and Cu-Mg) are explored.
Then from iterations 40 to 90, configurations of ternary alloys (Al-Cu-Mg) are sampled (see supporting information Table~\ref{table:S1} for more details). 
It is noted that in each iteration, $NpT$ simulations of 2 temperatures and 8 pressures covering the range $1\sim 50000$~bar are performed, thus configurations are sampled from 16 different thermodynamic conditions.
The temperatures increases as the iterations proceed, and the pressures are set to fixed values  (1, 10, 100, 1000, 5000, 10000, 20000, 50000~bar).
The surface initial configurations are generated by displacing two halves of FCC and HCP structures along certain crystallographic directions. 
The structures are prepared in two ways: (1) Conventional cells are copied along the displacing direction so that the thickness of surface structures is at least 10~\AA. 
(2) Conventional cells are replicated by $2\times 1 \times 1$ or $2 \times 2 \times 1$ times (see supporting information Table~\ref{table:S1} for the assignment of replications to crystals), and then are copied along the displacing direction so that the thickness of surface is at least three layers of atoms. 
Such sample sizes allow a balance between the rationality in simulating surface environments and the efficiency in successive labeling.
We chose \{100, 110, 111\} and \{001, 100, 110\} as directions of displacing for FCC and HCP, respectively.
The atomic positions of these displaced structures are randomly occupied by Al, Mg or Cu atoms. 
Simulation cells prepared by (1) are scaled with scaling factors from 0.96 to 1.06 with an interval of 0.02, and those prepared by (2) use scaling factors of 0.96, 1.00 and 1.06. 
Finally, the atomic positions and the cell shapes are randomly perturbed. 
Canonical ($NVT$) simulations are conducted to sample the surface configurations in the temperature range of $50.0 \sim 1222.0$~K. 
Binary and ternary alloys are explored from iteration 91 to 94, while Al-Cu and selected ternary alloys are explored from iteration 95 to 103 (see supporting information Table~\ref{table:S1}~for more information). 
The explored configurations are classified as accurate when the \emph{model deviation} is smaller than 0.05~eV/\AA, as candidate if the model deviation is between 0.05~eV/\AA\ and 0.20~eV/\AA, and as failed otherwise.

\paragraph{Labeling.}
The Vienna {\it ab initio} simulation package (VASP)\ucite{kresse1996efficient,kresse1996efficiency} is used to conduct the DFT calculation with the generalized gradient approximation (GGA) adopting Perdew-Burke-Ernzerhof (PBE)\ucite{Perdew1996PBE,PhysRevLett.78.1396}
exchange-correlation functional.
The projector-augmented-wave (PAW) method\textsuperscript{\cite{blochl1994projector,PhysRevB.59.1758}} is used with an energy cut-off of 650~eV for the plane wave basis sets.
K-points in the Brillouin zone  are sampled by the Monkhorst-Pack Mesh\textsuperscript{\cite{monkhorst1976special}} with grid spacing  of $h_k$ = 0.1~\AA$^{-1}$.
The order 1 Methfessel-Paxton smearing method\textsuperscript{\cite{methfessel1989high-precision}} is adopted with $\sigma$ = 0.22~eV. 
The convergence criterion for the self consistent field (SCF) calculation is $1 \times 10^{-6}$~eV for the changes of both the total energy and the band structure energy.

\paragraph{Training.}
The smooth edition of the Deep Potential model is adopted to construct the PES model.\textsuperscript{\cite{zhang2018end}} 
The DeePMD-kit package\textsuperscript{\cite{wang2018deepmd}} is used for training.
The sizes of the embedding and fitting nets are set to $(25, 50, 100)$ and $(240, 240, 240)$, respectively.
During the DP-GEN iterations, 
the cut-off radius is set to 6~\AA. 
The learning rate starts from $ 1\times10^{-3} $ and exponentially decays to $3.5\times10^{-8}$ after $ 1\times10^{6} $ training steps.
Four models are trained to construct the model ensemble. Using the same architecture and the same training data set, they only differ in the random seeds for initializing model parameters.
After the DP-GEN iterations being converged, the production models are trained with the cut-off radius set to 9~\AA\ and training steps set to $ 1.6\times10^{7}$.
In all training tasks, the Adam stochastic gradient descent method\textsuperscript{\cite{Kingma2015adam}} is used with default hyper-parameter settings provided by the TensorFlow package.\textsuperscript{\cite{abadi2015tensorflow:}}

\section{Results and discussion}\label{sec:results}
After the convergence of the DP-GEN process, about 2.73 billion alloy configurations are explored and only a small portion of them ($\sim0.0038\%$) are selected for labeling.
Data sets produced by this work have been uploaded to the online open data repository.\textsuperscript{\cite{DP-Lib_Al-Cu-Mg}}

In the following we systematically benchmark the DP model by testing its accuracy in terms of formation energy, equilibrium volume, equation of state (EOS), elastic modulus, surface formation energy, vacancy formation energy, and interstitial energies.
We consider the 58 alloy structures from the Materials Project (MP) database, including 15 Al-Cu, 30 Al-Mg, and 7 Cu-Mg binary structures, as well as 6 Al-Cu-Mg ternary structures.
The properties are calculated from the alloy structures relaxed at 0~bar.
A state-of-the-art MEAM\textsuperscript{\cite{jelinek2012modified}} model is used for comparison.
DFT calculations of these properties adopt the same numerical setup as those used in the labeling step. 
When computing the formation energy, equilibrium volume, interstitial energy as well as vacancy defect and surface formation energies, the ionic relaxation stops when all the forces in a configuration are smaller than $1 \times 10^{-2}$ eV/{\AA}. 
The elastic moduli are calculated by the strain-stress relation using the finite difference method, so it demands a tighter break criterion for both equilibrium and deformed structures. 
Here we terminate the ionic step when the energy change in two successive steps is smaller than  $1 \times 10^{-5}$~eV.
DP and MEAM ionic relaxations break when either one of the following two criteria is satisfied. 
(1) The energy change between two successive steps is smaller than one part in $1 \times 10^{12}$ of the total energy, or (2) the norm of the global force vector (could be approximately considered as the upper bound of any force component on any atom)
is smaller than $1 \times 10^{-6}$~eV/{\AA}.
All the computations and analyses of the properties mentioned above are facilitated by the automatic workflow implemented in DP-GEN,\textsuperscript{\cite{zhang2020dpgen}} which invokes VASP\textsuperscript{\cite{kresse1996efficient,kresse1996efficiency}} for DFT calculations and LAMMPS\textsuperscript{\cite{plimpton1995lammps}} for DP and MEAM calculations. 
The vacancy and interstitial structures are prepared by exhaustively generating all non-equivalent defect structures from the MP structures by the the python package pymatgen\textsuperscript{\cite{ong2012python}} (interstital structures are generated by the interstitialcy finding tool (InFit),\textsuperscript{\cite{10.3389/fmats.2017.00034}} which is integrated in the pymatgen package).   

\subsection{\label{sec:e-v}Formation energy and equilibrium volume}
The formation energy of binary alloys, taking Al-Cu for example, is defined as
\begin{eqnarray}
E^{\formation}_{\Al\Cu}  = E_{\Al\Cu}^{0}(c_{\Al})-c_{\Al}E^0_{\Al}-(1-c_{\Al})E^0_{\Cu}
\end{eqnarray}
and that of the ternary Al-Cu-Mg alloy is defined as
\begin{eqnarray}
E^{\formation}_{\Al\Cu\Mg} = E_{\Al\Cu\Mg}^{0}(c_{\Al},c_{\Cu})-c_{\Al}E^0_{\Al}-c_{\Cu}E^0_{\Cu}-(1-c_{\Al}-c_{\Cu})E^0_{\Mg}
\end{eqnarray}
where $E^{0}_{\Al\Cu}$ and $E^{0}_{\Al\Cu\Mg}$ denote the equilibrium energy per atom of the tested Al-Cu and Al-Cu-Mg alloys, respectively.
$c_\Al$ and $c_{\Cu}$ denote the component concentration  of $\Al$ and $\Cu$, respectively, in an alloy structure.
$E^0_{\Al}, E^0_{\Cu}$, and $E^0_{\Mg}$  denote the equilibrium energies per atom of the three pure metals in their stable crystalline structures.
Here Al and Cu take the FCC structure, and Mg takes the HCP structure.

\begin{figure}[htbp]  
    \centering   
        \subfloat%[] 
    {
        \begin{minipage}[t]{0.48\textwidth}
            \centering          
            \includegraphics[width=1.0\textwidth]{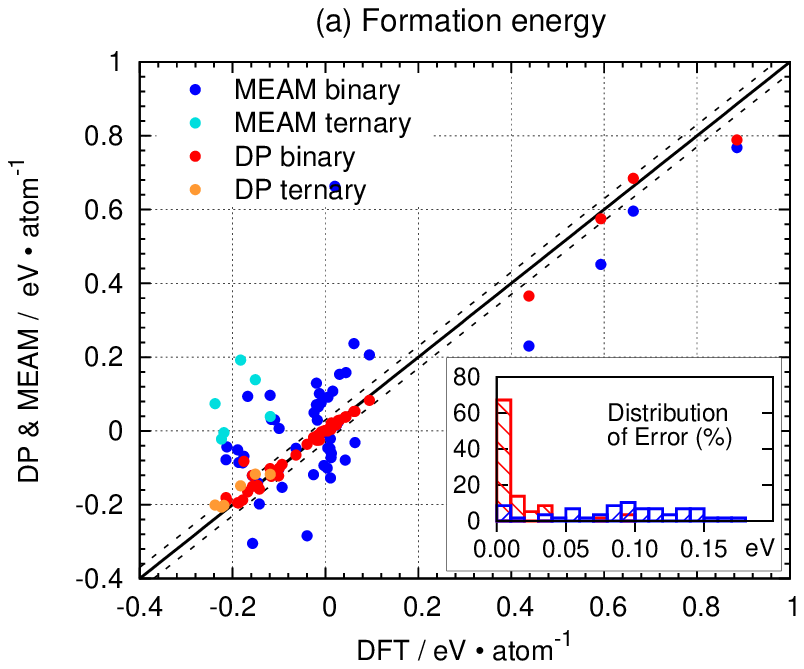} 
        \end{minipage}
    }
    \subfloat%[]
    {
        \begin{minipage}[t]{0.48\textwidth}
            \centering      
            \includegraphics[width=1.0\textwidth]{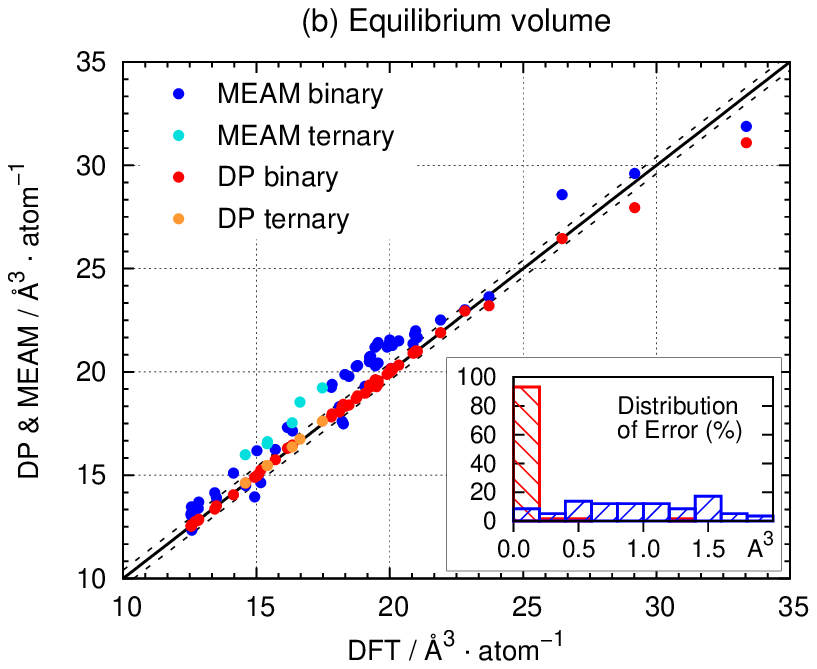}   
        \end{minipage}
    }
    \caption{Comparison of the formation energy and the equilibrium volume predicted by DP and MEAM for alloys, with respect to DFT results. (a) Formation energy with dashed lines representing $y = x \pm 0.03$ and the inserted histogram showing the absolute errors with respect to the DFT reference. (b) Equilibrium volume with dished lines representing $y = x \pm 0.2$ and the inserted histogram showing the distribution of errors. 58 structures, including both binary and ternary alloy structures, are adopted from Materials Projects (MP) database.}
     \label{fig:equi}
    % \WR {data in Fig. 1(a) revised}
\end{figure}

The DP predictions of formation energy and equilibrium volume for all tested structures are in satisfactory consistency with DFT results. 
In comparison with MEAM results, the relative errors with respect to DFT references are generally smaller.
As shown by Fig.~\ref{fig:equi}(a), the absolute differences between DP and DFT formation energies are below 10~meV${\cdot}$atom$^{-1}$ for 39 MP structures (67\% in total 58 MP structures), and no tested case has an error larger than 100~meV${\cdot}$atom$^{-1}$ 
As shown in Fig.~\ref{fig:equi}(b), 54 MP structures, or 93\% of the tested structures, have absolute errors in equilibrium volumes of < 0.2~\AA$^3{\cdot}$atom$^{-1}$. 
Compared with MEAM results, the errors of most of the testing cases in the formation energy (in~eV${\cdot}$atom$^{-1}$)
are reduced by at least one order of magnitude, and the relative errors in equilibrium volume are generally reduced from several percents to beneath 1\%. 
About 53\% of the MEAM results give errors over 100~meV${\cdot}$atom$^{-1}$. 
About 17\% in total give errors larger than 200~meV${\cdot}$
atom$^{-1}$, so they are beyond the plotted range of the distribution of errors.

\subsection{\label{sec:eos}Equation of state}
\begin{figure}
\begin{center}
\includegraphics[width=0.98\textwidth]{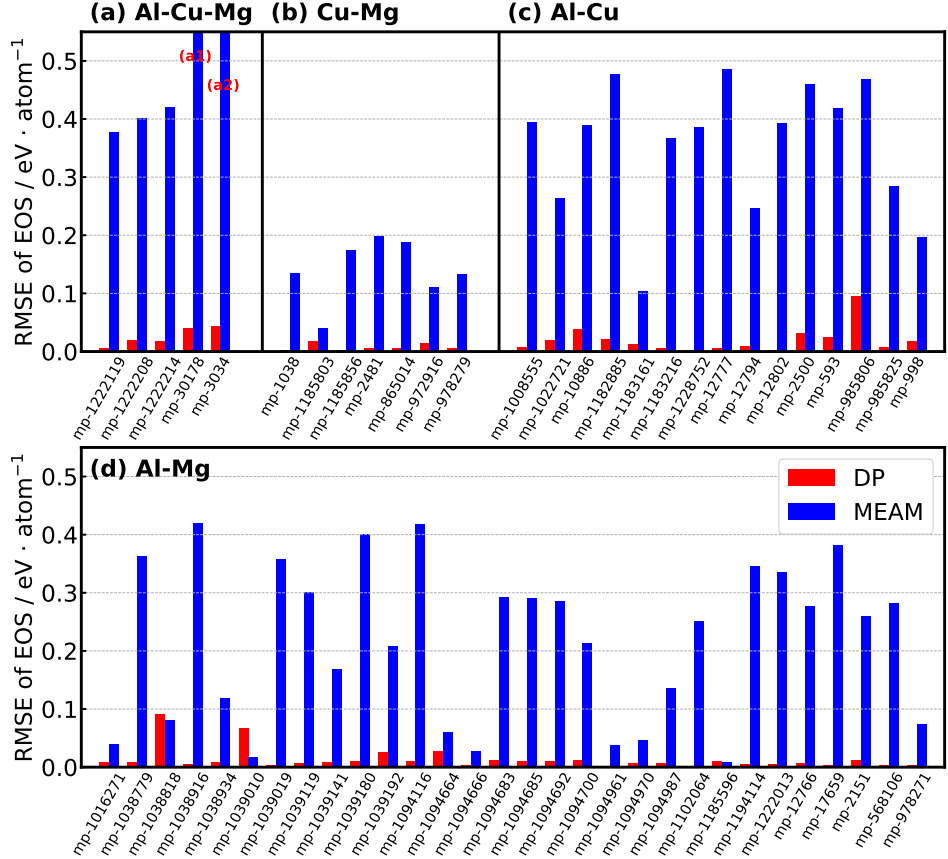}\\[5pt]  % insert figure
\caption{Comparison of DP and MEAM root-mean-squared error (RMSE) of EOS for alloys, with respect to DFT results. Structures are downloaded from the materials MP database. Volume range for computing RMSE is approximately from 80\% to 120\% of the DFT equilibrium volume. (a) 5 Al-Cu-Mg structures, 2 of them give MEAM errors larger than the plotting range: (a1) 0.600~eV${\cdot}$atom$^{-1}$; (a2) 0.663~eV${\cdot}$atom$^{-1}$. MP-1200279 is not included due to the convergence difficulty in the DFT calculation; (b) 7 Cu-Mg structures; (c) 15 Al-Cu structures. (d) 30 Al-Mg structures. The DP and MEAM EOS of MP-1039192 are estimated by single-point energy calculations on converged structures from the DFT EOS computation, because the MP-1039192 structure is not stable in neither DP nor MEAM energy minimization.
}\label{fig:eos}
\end{center}
\end{figure}

Next, we investigate the energy dependence on the volume region around the equilibrium value ($80 \%\sim 120\%$ of the equilibrium volume). 
The performance of DP on the equation of state (EOS) for both binary and ternary alloys are shown in Fig.~\ref{fig:eos}. 
The absolute DP root-mean-squared errors (RMSEs) with respect to DFT references are smaller than 10~meV${\cdot}$atom$^{-1}$ for 34 out of 57 tested alloys (MP-1200279 is not included due to the convergence difficulty in the DFT calculation), which is close to the performance on equilibrium energy (39 out of 58 structures with errors smaller than 10~meV${\cdot}$atom$^{-1}$). 
This indicates that DP model generally holds its accuracy for the equilibrium state in this volume range.
As a comparison, for more than half of the tested cases, the RMSEs of DP predictions are two orders of magnitudes smaller than MEAM results, thus DP ensures the generally more reliable energy-volume dependence than MEAM for crystals in the whole Al-Cu-Mg concentration space.

We note that, for all the properties introduced above, despite an overall good agreement with DFT values, DP gives unsatisfactory predictions on a small number of structures.
A detailed investigation suggests that this may be attributed to the exploration strategy adopted by the DP-GEN process.
The MD simulation for exploring the conformation space starts from structures with FCC, HCP, and BCC lattices, and should include the corresponding crystalline, molten, and surface structures.
However, there are local chemical environments that can be hardly sampled using this strategy, so that DP will be in the extrapolation regime when tested on the associated structures.
Examples include MP-1038818 and MP-1039010 in Fig.~\ref{fig:eos}, which are in diamond and simple cubic lattices, respectively.
It should be mentioned that, however, if we continue with the DP-GEN process and add related structures into the exploration step, the DP model should be systematically improved.
As an evidence, we refer to Ref.\ucite{zhang2019active} for an Al-Mg model generated by a DP-GEN process that takes into considerations the diamond and simple cubic structures.
Such a model shows a much better prediction accuracy on MP-1038818 and MP-1039010.

\subsection{Elastic modulus and surface formation energy}

Now we consider the bulk and shear moduli as well as the unrelaxed surface formation energy. The unrelaxed surface formation energy, $E^{sf}((lmn))$, refers to the energy needed to create a surface with the miller index $(lmn)$ for a given crystal, and is defined as 
\begin{eqnarray}
E^{sf}((lmn)) = \frac{1}{2A} (E^{s}((lmn)) - N^sE^0)
\label{eq:surf}
\end{eqnarray}
Here $E^{s}((lmn))$ and $N^{s}$ denotes the energy and number of atoms of the unrelaxed surface structure with the Miller index $(lmn)$. 
$A$ denotes the surface area. For each MP case, $E^0$ refers to the per-atom potential energy of the relaxed bulk structure (under 0~bar) of this MP crystal. For each surface, energy evaluations of $E^{s}((lmn))$ using three methods are conducted on the same surface geometry, i.e., the one generated by displacing two halves of the DFT equilibrium bulk structure. 
This ensures the rationality of the comparison, since a small difference in the equilibrium bulk structure relaxed by different methods and codes could bring distinct surface geometries under the same Miller index $(lmn)$ (surface generation methods is similar to that described in Sec.~\ref{sec:method}~~~).

\begin{figure}[htbp]
    \centering   
    
    \subfloat%[] 
    {
        \begin{minipage}[t]{0.48\textwidth}
            \centering          
            \includegraphics[width=1.0\textwidth]{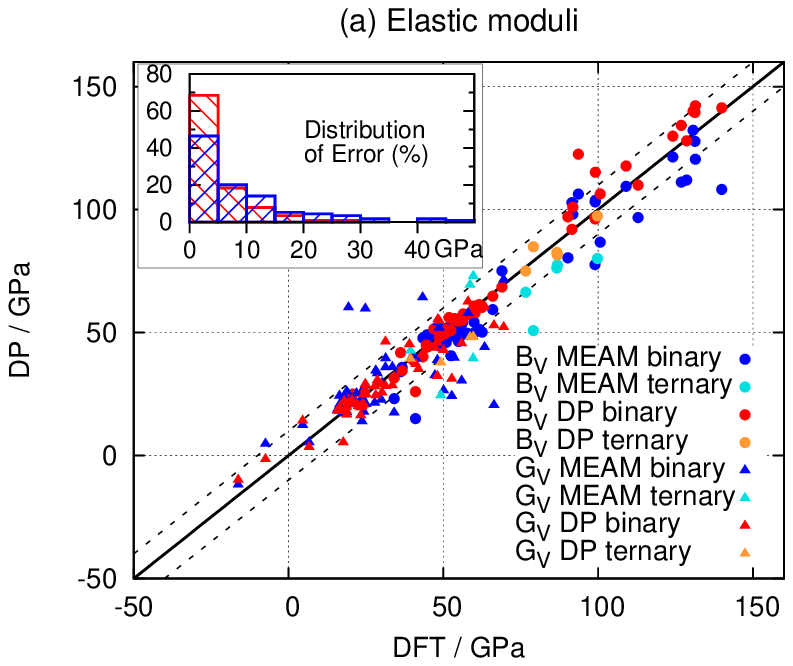} 
        \end{minipage}
    }
    \subfloat%[]
    {
        \begin{minipage}[t]{0.48\textwidth}
            \centering      
            \includegraphics[width=1.0\textwidth]{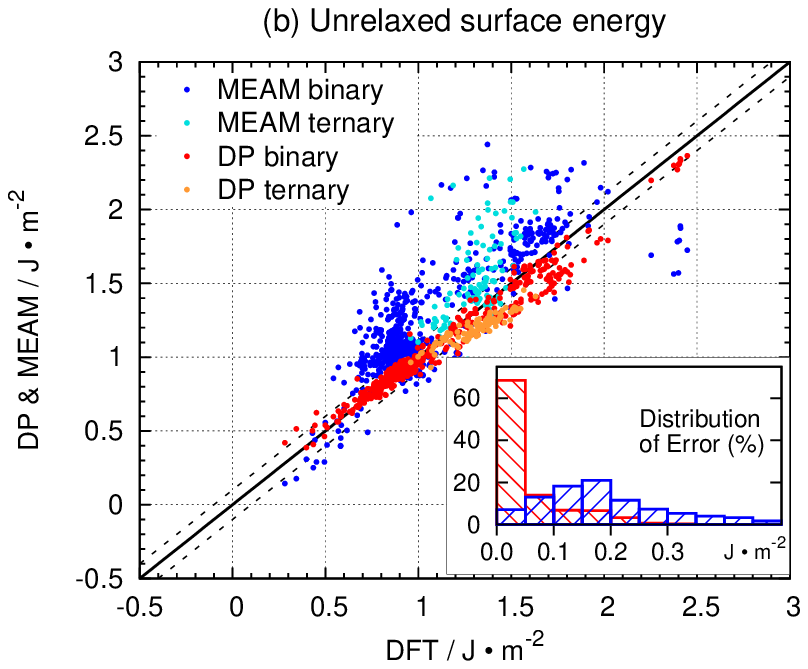}   
        \end{minipage}
    }
    
    \caption{Comparison of the elastic moduli and the unrelaxed surface formation energy predicted by DP and MEAM for alloys, with respect to DFT results. (a) The bulk modulus $B_v$ and the shear modulus $G_v$ of 57 MP structures, with dashed lines representing $y = x \pm 10$ 
    and the inserted histogram showing the distribution of absolute errors with respect to the DFT reference. MP-1200279 is not included due to the convergence difficulty in the DFT calculation. (b) 911 unrelaxed surface formation energies of 56 MP structures, with dashed lines representing $y = x \pm 0.1$ and the inserted histogram showing distribution of errors. MP-1200279 and MP-1185596 are not included due to the convergence difficulty in the DFT calculation.}
    \label{fig:elastic-surf}
\end{figure}

As shown in Fig.~\ref{fig:elastic-surf}(a), DP generally gives more accurate bulk and shear moduli than MEAM. It presents a more concentrated distribution of data points along the diagonal line denoting the DFT reference.
This advantage is more directly revealed by comparing the absolute errors with respect to the DFT reference. Within the leftmost interval of the inserted histogram that allows errors beneath 5~GPa, about 20\% more results in total are presented by DP than MEAM. Meanwhile, the tail of the distribution of error is also shorter for DP. 
Furthermore, both bulk and shear moduli of all 5 tested ternary cases are more accurately predicted by DP.

DP also outperforms MEAM on predictions of unrelaxed surface formation energy, as shown by Fig.~\ref{fig:elastic-surf}(b). 
Taking DFT results as references and compared with MEAM, there are 50\% more data points with absolute errors smaller than 0.05~$J/m^{2}$ when using DP for predictions.
However, DP slightly underestimates $E^{sf}$ for some ternary cases and a small portion of binary cases. 

\subsection{Vacancy and interstitial defects}

Finally, for ternary alloys, we examine the vacancy defect formation energy and the energy along interstitial relaxation paths. 
The vacancy formation energy $E^{vf}$ is defined as 
\begin{eqnarray}
E^{vf} = E^{v}(N^{v}) - E^{0}N^{v}
\end{eqnarray}
where $E^{v}(N^{v})$ denotes the energy of a relaxed structure with one vacancy defect and $N^{v}$ atoms. $E^{0}$ is for equilibrium 
energy per atom of the corresponding pristine crystal.

For the interstitial defect, due to the generally high energetic instability of the generated interstitial structures, the relaxation may not end up with structures with locally relaxed interstitial point defects. Meanwhile, the relaxation path would be highly sensitive to the practical implementation of the relaxation algorithm.\textsuperscript{\cite{zhang2019active}}
Therefore, instead of comparing the interstitial formation energies of DFT, DP and MEAM calculated by independent relaxations, we investigate the DP and the MEAM prediction errors along the DFT relaxation path (excluding early high energy geometries).

\begin{figure}[htbp]  
    \centering   
    
    \subfloat%[Vacancy formation energy]
    {
        \begin{minipage}[t]{0.48\textwidth}
            \centering        
            \includegraphics[width=1.0\textwidth]{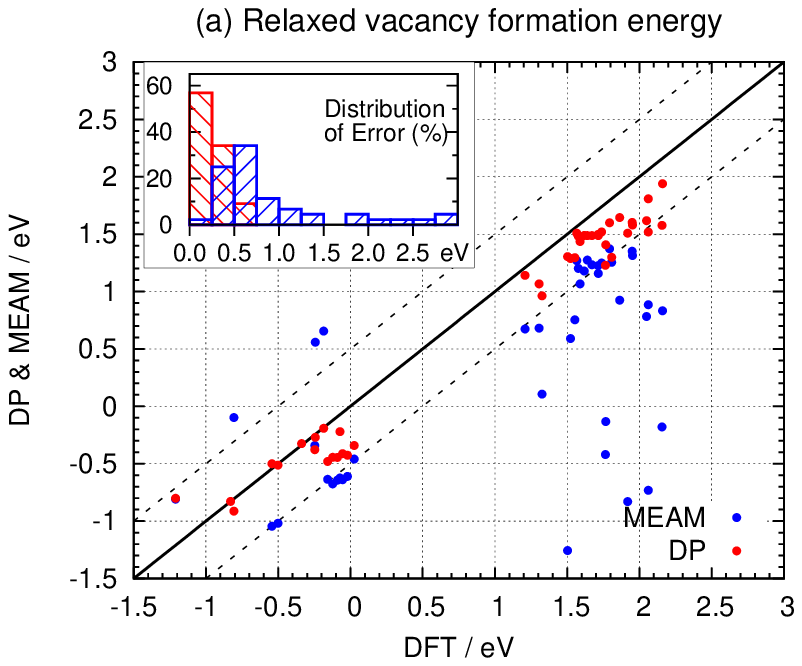}   
        \end{minipage}
    }
    \subfloat%[Energy along interstitial relaxation path]
    {
        \begin{minipage}[t]{0.48\textwidth}
            \centering     
            \includegraphics[width=1.0\textwidth]{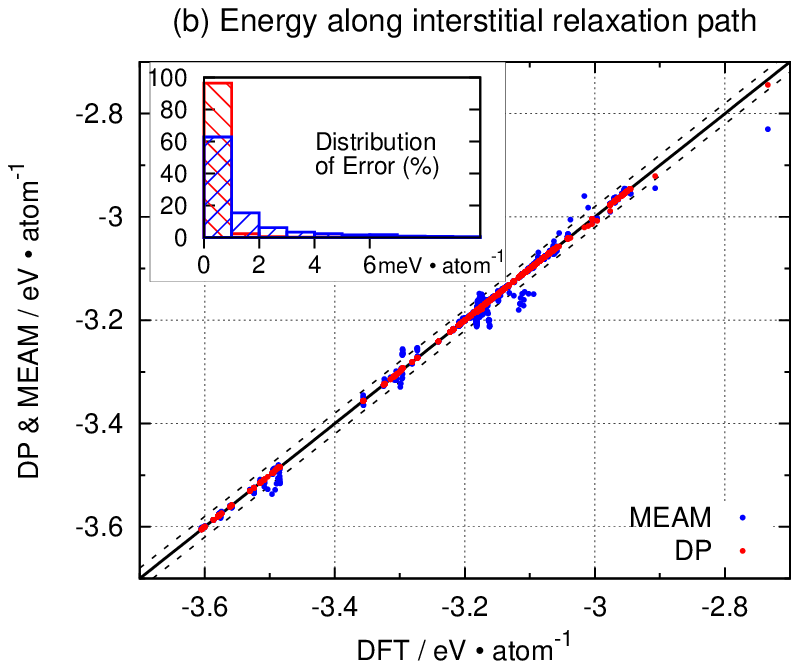}   
        \end{minipage}
    }
    
    \caption{Comparison of the vacancy formation energy and the unrelaxed surface formation energy predicted by DP and MEAM for ternary alloys, with respect to DFT results. (a) 44 relaxed vacancy formation energies of 6 ternary alloys, with dashed lines representing $ y = x \pm 0.5$ and the inserted panel showing the error distribution. (b) 23168 energies per atom along DFT relaxation path of 5 ternary alloys, with dashed lines representing $ y = x \pm 0.02$ and the inserted panel showing the error distribution. MP-1200279 is not included due to the convergence difficulty in the DFT calculation.} 
    \label{fig:defect}
\end{figure}

DP is in satisfactory consistency with DFT and shows relative advantages compared with MEAM for both properties.
As shown by Fig.~\ref{fig:defect}(a), almost all DP predicted $E^{vf}$ values give absolute errors beneath 0.5~eV with respect to DFT values. 
It can be found that, DP tends to slightly underestimate $E^{vf}$ compared with the DFT reference.
In terms of the energy along DFT interstitial relaxation paths, nearly all DP predictions give absolute errors beneath 1 meV${\cdot}$atom$^{-1}$. 
In contrast, MEAM gives about 30\% less data points in total within the same error range. 
Meanwhile,  
some MEAM predictions have errors of tens of meV${\cdot}$atom$^{-1}$, as demonstrated by those points far from the DFT reference line in Fig.~\ref{fig:defect}(b).

It is noted that the vacancy and interstitial formation energies calculated by the tested structures may not be well converged due to the finite size effect. 
However, for the purpose of investigating the accuracy of the DP and the MEAM in terms of the defect formation energy, it is reasonable to compare the results of DP and MEAM with that of the DFT, computed from the defect structures with the same number of atoms. 

\section{Conclusion}\label{sec:conclusion}

Using the Deep Potential representation and the concurrent-learning scheme, DP-GEN, we develop an accurate PES model for the Al-Cu-Mg alloys in the full concentration range. 
Systematic benchmarks with DFT results on fundamental energetic, mechanical, and defect properties of 58 MP structures suggest a good accuracy in the full concentration space, and an overall better performance than a state-of-the-art MEAM potential.
The training does not rely on any existed first-principles data set. 
Instead, data is generated on-the-fly through DP-GEN and finally form a compact sets supporting reliable DP model after the scheme is converged. 
The resulting DP model and data sets are expected to help investigate the physical mechanisms behind complex phenomena of Al-Cu-Mg systems and contribute to the methodology development for materials simulation.

\addcontentsline{toc}{chapter}{Acknowledgment}
\section*{Acknowledgment}
The authors are grateful for the computing time provided by the High-performance Computing Platform of Peking University, the Beijing Institute of Big Data Research and the Alibaba cloud.
The work of H.W. is supported by the National Science Foundation of China under Grant No.~11871110, the National Key Research and Development Program of China under Grants No.~2016YFB0201200 and No.~2016YFB0201203, and Beijing Academy of Artificial Intelligence (BAAI).

%\end{CJK*}  %% end the Chinese environment
\newpage

\section*{Supporting Information}

\vspace{5mm}
\renewcommand\thefigure{S\arabic{figure}}
\setcounter{figure}{0}
\setcounter{section}{0}
\let\cleardoublepage\clearpage
\section{Supplementary Text}

\begin{figure}[htbp]
    \centering   
        \subfloat%[] 
    {
        \begin{minipage}[t]{0.68\textwidth}
            \centering
            \includegraphics[width=1.0\textwidth]{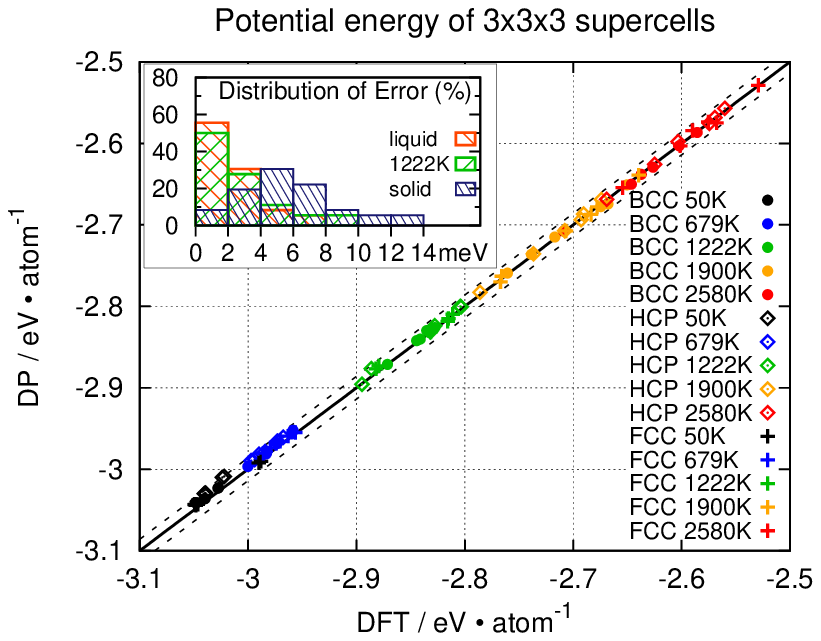} 
        \end{minipage}
         
    }
   \caption{Comparison of the potential energy predicted by DP with those calculated by DFT for equimolar ternary alloys in $3\times 3\times 3$ super-cells. 90 configurations are investigated. They are taken from 45 isothermo-isobaric DPMD trajectories at 300~ps and 350~ps. The 45 DPMD trajectories are simulated at 5 temperatures (about 50~K, 679~K, 1222~K, 1900~K and 2580~K) and 3 pressures (1~bar, 5000~bar and 50000~bar), starting from 3 kinds of equimolar structures respectively in BCC, HCP and FCC lattices. The occupation of atomic species on corresponding lattice positions in the initial structures is randomly assigned for each trajectory. Dashed lines represents $y = x \pm 0.014$. The inserted histogram shows the distribution of errors, with configurations from 50~K and 679~K trajectories being assigned to the solid state and those from 1900~K and 2580~K trajectories to the liquid state.}
     \label{fig:Potentia energy 333}
\end{figure}

\begin{figure}[htbp]
    \centering   
        \subfloat%[] 
    {
        \begin{minipage}[t]{0.68\textwidth}
            \centering          
            \includegraphics[width=1.0\textwidth]{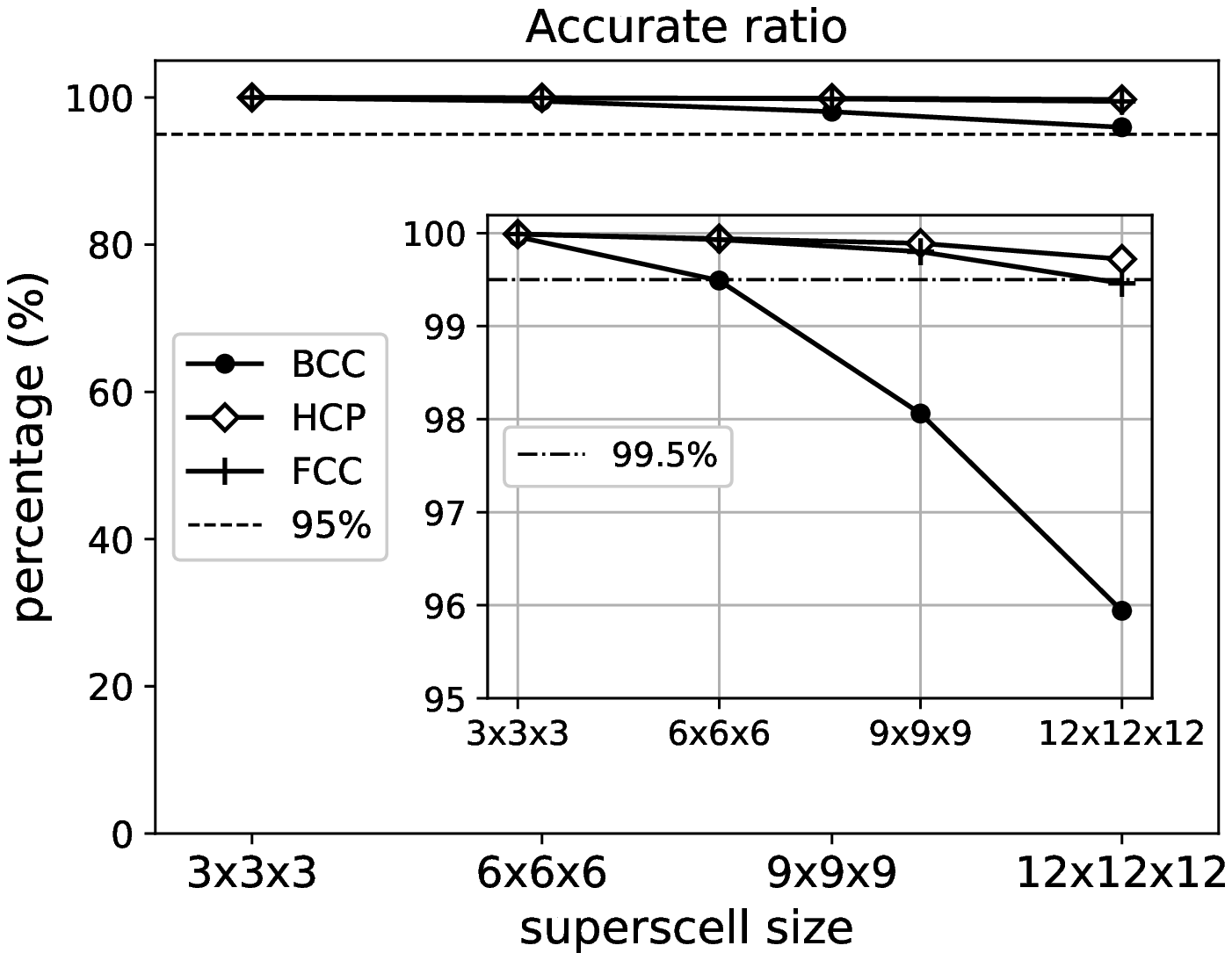} 
        \end{minipage}
    }
    \caption{Accurate ratios on larger equimolar ternary alloys of DP models generated by DP-GEN after the convergence of the workflow. 
    The prediction of the model is classified as "accurate" if the model deviation is smaller than 0.05~eV/\AA.
    We consider the accurate ratios of configurations in 4 super-cell sizes for BCC, HCP and FCC lattices.
    For each super-cell, 480 DPMD simulations are conducted for 3 lattices at 20 temperatures approximately from 50~K to 2580~K and 8 pressures from 1~bar to 50000~bar.
    Each DPMD lasts for 6~ps, and configurations are sampled every 20~fs.
    In total 144000 configurations are sampled for each super-cell at all thermodynamic conditions.
    The occupation of different elements on the lattice points is randomly assigned for each initial configurations.}
     \label{fig:accu ratio}
\end{figure}

DP model treats the potential energy of a system as the summation of the energy contribution from each atom, and predicts the energy of each atom according to its local atomic environment.
$2\times 2\times 2$ super-cells, used to construct the training sets, is large enough to capture the local atomic environments needed for generating an accurate DP model for Al-Cu-Mg alloys.
Verification is conducted by using corresponding DP models to predict alloy structures with larger super-cells including $3\times 3\times 3$, $6\times 6\times 6$, $9\times 9\times 9$ and $12\times 12\times 12$.
The example system are ternary alloy in the equimolar concentration.
With tolerable DFT computational cost, for $3\times 3\times 3$ super-cells, the predicted potential energy are directly compared with DFT results.
As shown by Fig.~\ref{fig:Potentia energy 333}, above 90\% cases give relative errors smaller than 10~meV/atom and no case gives the error above 14~meV/atom.
For even larger super-cells where computational cost of DFT becomes more impractical, the model deviation defined in DP-GEN scheme is used, that avoids using expensive DFT labels as references.
As shown by the Fig.~\ref{fig:accu ratio}, the accuracy keeps higher than 99.5\% for all configurations till $6\times 6\times 6$ and is still higher than 95\% when using $12\times 12\times 12$ super-cells.

\newpage
\section{Supplementary Figures}

\newpage

\section{Supplementary Table}

\begin{center}
\begin{table}[H]
\setlength{\abovecaptionskip}{0.cm}
\setlength{\belowcaptionskip}{-0.9cm}
\renewcommand{\thetable}{S\arabic{table}}
\caption{Details of the exploration strategy for the Al-Cu-Mg alloys.}\label{table:S1}
\resizebox{\textwidth}{!}{
\begin{threeparttable}
\begin{tabular}{cccccccc}
\hline
Iter& Alloy &Crystal&DPMD steps&length/ps&Temperature/K&Ensemble&Candidates per/\%\\
\hline
0& binary & FCC,HCP,BCC & 500 & 1 & 50, 135.7 & $NpT$\tnote{a} & 15.08 \\
1& binary & FCC,HCP,BCC & 1000 & 2 & 50, 135.7 & $NpT$\tnote{a} & 0.71 \\
2& binary & FCC,HCP,BCC & 3000 & 6 & 50, 135.7 & $NpT$\tnote{a} & 0.12 \\
3& binary & FCC,HCP,BCC & 3000 & 6 & 50, 135.7 & $NpT$\tnote{a} & 0.20 \\
4& binary & FCC,HCP,BCC & 500 & 1 & 272.6, 407.3 & $NpT$\tnote{a} & 2.21 \\
5& binary & FCC,HCP,BCC & 1000 & 2 & 272.6, 407.3 & $NpT$\tnote{a} & 1.20 \\
6& binary & FCC,HCP,BCC & 3000 & 6 & 272.6, 407.3 & $NpT$\tnote{a} & 0.17 \\
7& binary & FCC,HCP,BCC & 3000 & 6 & 272.6, 407.3 & $NpT$\tnote{a} & 0.05 \\
8& binary & FCC,HCP,BCC & 500 & 1 & 543.1, 678.9 & $NpT$\tnote{a} & 1.04 \\
9& binary & FCC,HCP,BCC & 1000 & 2 & 543.1, 678.9 & $NpT$\tnote{a} & 0.93 \\
10& binary & FCC,HCP,BCC & 3000 & 6 & 543.1, 678.9 & $NpT$\tnote{a} & 0.11 \\
11& binary & FCC,HCP,BCC & 3000 & 6 & 543.1, 678.9 & $NpT$\tnote{a} & 0.05 \\
12& binary & FCC,HCP,BCC & 500 & 1 & 814.7, 950.4 & $NpT$\tnote{a} & 0.99 \\
13& binary & FCC,HCP,BCC & 1000 & 2 & 814.7, 950.4 & $NpT$\tnote{a} & 0.94 \\
14& binary & FCC,HCP,BCC & 3000 & 6 & 814.7, 950.4 & $NpT$\tnote{a} & 0.15 \\
15& binary & FCC,HCP,BCC & 3000 & 6 & 814.7, 950.4 & $NpT$\tnote{a} & 0.09 \\
16& binary & FCC,HCP,BCC & 500 & 1 & 1086.2, 1222.0 & $NpT$\tnote{a} & 1.00 \\
17& binary & FCC,HCP,BCC & 1000 & 2 & 1086.2, 1222.0 & $NpT$\tnote{a} & 0.80 \\
18& binary & FCC,HCP,BCC & 3000 & 6 & 1086.2, 1222.0 & $NpT$\tnote{a} & 0.10 \\
19& binary & FCC,HCP,BCC & 3000 & 6 & 1086.2, 1222.0 & $NpT$\tnote{a} & 0.12 \\
20& binary & FCC,HCP,BCC & 500 & 1 & 1357.8, 1493.5 & $NpT$\tnote{a} & 1.15 \\
21& binary & FCC,HCP,BCC & 1000 & 2 & 1357.8, 1493.5 & $NpT$\tnote{a} & 0.42 \\
22& binary & FCC,HCP,BCC & 3000 & 6 & 1357.8, 1493.5 & $NpT$\tnote{a} & 0.17 \\
23& binary & FCC,HCP,BCC & 3000 & 6 & 1357.8, 1493.5 & $NpT$\tnote{a} & 0.27 \\
24& binary & FCC,HCP,BCC & 500 & 1 & 1629.3, 1765.1 & $NpT$\tnote{a} & 0.74 \\
25& binary & FCC,HCP,BCC & 1000 & 2 & 1629.3, 1765.1 & $NpT$\tnote{a} & 0.30 \\
26& binary & FCC,HCP,BCC & 3000 & 6 & 1629.3, 1765.1 & $NpT$\tnote{a} & 0.22 \\
27& binary & FCC,HCP,BCC & 3000 & 6 & 1629.3, 1765.1 & $NpT$\tnote{a} & 0.23 \\
28& binary & FCC,HCP,BCC & 500 & 1 & 1900.9, 2036.7 & $NpT$\tnote{a} & 0.88 \\
29& binary & FCC,HCP,BCC & 1000 & 2 & 1900.9, 2036.7 & $NpT$\tnote{a} & 0.94 \\
30& binary & FCC,HCP,BCC & 3000 & 6 & 1900.9, 2036.7 & $NpT$\tnote{a} & 0.21 \\
31& binary & FCC,HCP,BCC & 3000 & 6 & 1900.9, 2036.7 & $NpT$\tnote{a} & 0.23 \\
\hline
\end{tabular}
\end{threeparttable}}
\end{table}
\end{center}

\begin{center}
\begin{table}[H]
\resizebox{\textwidth}{!}{
\begin{threeparttable}
\begin{tabular}{cccccccc}
\hline
Iter& Alloy & Crystal & DPMD steps & length/ps & T/K & Ensemble & Candidates per/\% \\
\hline
32& binary & FCC,HCP,BCC & 500 & 1 & 2172.4,2308.2 & $NpT$\tnote{a} & 0.81 \\
33& binary & FCC,HCP,BCC & 1000 & 2 & 2172.4, 2308.2 & $NpT$\tnote{a} & 0.27 \\
34& binary & FCC,HCP,BCC & 3000 & 6 & 2172.4, 2308.2 & $NpT$\tnote{a} & 0.25 \\
35& binary & FCC,HCP,BCC & 3000 & 6 & 2172.4, 2308.2 & $NpT$\tnote{a} & 0.13 \\
36& binary & FCC,HCP,BCC & 500 & 1 & 2444.0, 2579.8 & $NpT$\tnote{a} & 0.60 \\
37& binary & FCC,HCP,BCC & 1000 & 2 & 2444.0, 2579.8 & $NpT$\tnote{a} & 0.41 \\
38& binary & FCC,HCP,BCC & 3000 & 6 & 2444.0, 2579.8 & $NpT$\tnote{a} & 0.27 \\
39& binary & FCC,HCP,BCC & 3000 & 6 & 2444.0, 2579.8 & $NpT$\tnote{a} & 0.48 \\
40& ternary & FCC,HCP,BCC & 500 & 1 & 50, 135.7 & $NpT$\tnote{a} & 77.68 \\
41& ternary & FCC,HCP,BCC & 1000 & 2 & 50, 135.7 & $NpT$\tnote{a} & 0.12 \\
42& ternary & FCC,HCP,BCC & 3000 & 6 & 50, 135.7 & $NpT$\tnote{a} & 0.01 \\
43& ternary & FCC,HCP,BCC & 3000 & 6 & 50, 135.7 & $NpT$\tnote{a} & 0.00 \\
44& ternary & FCC,HCP,BCC & 500 & 1 & 272.6, 407.3 & $NpT$\tnote{a} & 0.05 \\
45& ternary & FCC,HCP,BCC & 1000 & 2 & 272.6, 407.3 & $NpT$\tnote{a} & 0.01 \\
46& ternary & FCC,HCP,BCC & 3000 & 6 & 272.6, 407.3 & $NpT$\tnote{a} & 0.01 \\
47& ternary & FCC,HCP,BCC & 3000 & 6 & 272.6, 407.3 & $NpT$\tnote{a} & 0.01 \\
48& ternary & FCC,HCP,BCC & 500 & 1 & 543.1, 678.9 & $NpT$\tnote{a} & 0.07 \\
49& ternary & FCC,HCP,BCC & 1000 & 2 & 543.1, 678.9 & $NpT$\tnote{a} & 0.04 \\
50& ternary & FCC,HCP,BCC & 3000 & 6 & 543.1, 678.9 & $NpT$\tnote{a} & 0.00 \\
51& ternary & FCC,HCP,BCC & 3000 & 6 & 543.1, 678.9 & $NpT$\tnote{a} & 0.00 \\
52& ternary & FCC,HCP,BCC & 500 & 1 & 814.7, 950.4 & $NpT$\tnote{a} & 0.05 \\
53& ternary & FCC,HCP,BCC & 1000 & 2 & 814.7, 950.4 & $NpT$\tnote{a} & 0.02 \\
54& ternary & FCC,HCP,BCC & 3000 & 6 & 814.7, 950.4 & $NpT$\tnote{a} & 0.01 \\
55& ternary & FCC,HCP,BCC & 3000 & 6 & 814.7, 950.4 & $NpT$\tnote{a} & 0.01 \\
56& ternary & FCC,HCP,BCC & 500 & 1 & 1086.2, 1222.0 & $NpT$\tnote{a} & 0.03 \\
57& ternary & FCC,HCP,BCC & 1000 & 2 & 1086.2, 1222.0 & $NpT$\tnote{a} & 0.04 \\
58& ternary & FCC,HCP,BCC & 3000 & 6 & 1086.2, 1222.0 & $NpT$\tnote{a} & 0.01 \\
59& ternary & FCC,HCP,BCC & 3000 & 6 & 1086.2, 1222.0 & $NpT$\tnote{a} & 0.01 \\
60& ternary & FCC,HCP,BCC & 500 & 1 & 1357.8, 1493.5 & $NpT$\tnote{a} & 0.16 \\
61& ternary & FCC,HCP,BCC & 1000 & 2 & 1357.8, 1493.5 & $NpT$\tnote{a} & 0.03 \\
62& ternary & FCC,HCP,BCC & 3000 & 6 & 1357.8, 1493.5 & $NpT$\tnote{a} & 0.02 \\
63& ternary & FCC,HCP,BCC & 3000 & 6 & 1357.8, 1493.5 & $NpT$\tnote{a} & 0.02 \\
\hline
\end{tabular}
\end{threeparttable}}
\end{table}
\end{center}

\begin{center}
\begin{table}[H]
\resizebox{\textwidth}{!}{
\begin{threeparttable}
\begin{tabular}{cccccccc}
\hline
Iter& Alloy & Crystal & DPMD steps & length/ps & T/K & Ensemble & Candidates per/\% \\
\hline
64& ternary & FCC,HCP,BCC & 500 & 1 & 1629.3, 1765.1 & $NpT$\tnote{a} & 0.07 \\
65& ternary & FCC,HCP,BCC & 1000 & 2 & 1629.3, 1765.1 & $NpT$\tnote{a} & 0.04 \\
66& ternary & FCC,HCP,BCC & 3000 & 6 & 1629.3, 1765.1 & $NpT$\tnote{a} & 0.01 \\
67& ternary & FCC,HCP,BCC & 3000 & 6 & 1629.3, 1765.1 & $NpT$\tnote{a} & 0.02 \\
68& ternary & FCC,HCP,BCC & 500 & 1 & 1900.9, 2036.7 & $NpT$\tnote{a} & 0.08 \\
69& ternary & FCC,HCP,BCC & 1000 & 2 & 1900.9, 2036.7 & $NpT$\tnote{a} & 0.02 \\
70& ternary & FCC,HCP,BCC & 3000 & 6 & 1900.9, 2036.7 & $NpT$\tnote{a} & 0.02 \\
71& ternary & FCC,HCP,BCC & 3000 & 6 & 1900.9, 2036.7 & $NpT$\tnote{a} & 0.01 \\
72& ternary & FCC,HCP,BCC & 500 & 1 & 2172.4,2308.2 & $NpT$\tnote{a} & 0.07 \\
73& ternary & FCC,HCP,BCC & 1000 & 2 & 2172.4, 2308.2 & $NpT$\tnote{a} & 0.10 \\
74& ternary & FCC,HCP,BCC & 3000 & 6 & 2172.4, 2308.2 & $NpT$\tnote{a} & 0.03 \\
75& ternary & FCC,HCP,BCC & 3000 & 6 & 2172.4, 2308.2 & $NpT$\tnote{a} & 0.03 \\
76& ternary & FCC,HCP,BCC & 500 & 1 & 2444.0, 2579.8 & $NpT$\tnote{a} & 0.09 \\
77& ternary & FCC,HCP,BCC & 1000 & 2 & 2444.0, 2579.8 & $NpT$\tnote{a} & 0.05 \\
78& ternary & FCC,HCP,BCC & 3000 & 6 & 2444.0, 2579.8 & $NpT$\tnote{a} & 0.05 \\
79& ternary & FCC,HCP,BCC & 3000 & 6 & 2444.0, 2579.8 & $NpT$\tnote{a} & 0.02 \\
80& ternary & FCC,HCP,BCC & 3000 & 6 & 50, 135.7 & $NpT$\tnote{b} & 0.00 \\
81& ternary & FCC,HCP,BCC & 6000 & 12 & 50, 135.7 & $NpT$\tnote{b} & 0.00 \\
82& ternary & FCC,HCP,BCC & 3000 & 6 & 272.6, 407.3 & $NpT$\tnote{b} & 0.00 \\
83& ternary & FCC,HCP,BCC & 6000 & 12 & 272.6, 407.3 & $NpT$\tnote{b} & 0.00 \\
84& ternary & FCC,HCP,BCC & 3000 & 6 & 543.1, 678.9 & $NpT$\tnote{b} & 0.00 \\
85& ternary & FCC,HCP,BCC & 6000 & 12 & 543.1, 678.9 & $NpT$\tnote{b} & 0.00 \\
86& ternary & FCC,HCP,BCC & 3000 & 6 & 814.7, 950.4 & $NpT$\tnote{b} & 0.00 \\
87& ternary & FCC,HCP,BCC & 6000 & 12 & 814.7, 950.4 & $NpT$\tnote{b} & 0.00 \\
88& ternary & FCC,HCP,BCC & 3000 & 6 & 1086.2, 1222.0 & $NpT$\tnote{b} & 0.00 \\
89& ternary & FCC,HCP,BCC & 6000 & 12 & 1086.2, 1222.0 & $NpT$\tnote{b} & 0.00 \\
90& ternary & FCC,HCP,BCC & 3000 & 6 & 1357.8, 1493.5 & $NpT$\tnote{b} & 0.00 \\
91& binary\&ternary & FCC(surf)\tnote{c}, HCP(surf)\tnote{c} & 1000 & 2 & 50, 135.7 & $NVT$ & 40.16 \\
92& binary\&ternary & FCC(surf)\tnote{c}, HCP(surf)\tnote{c} & 3000 & 6 & 50, 135.7 & $NVT$ & 0.06 \\
93& binary\&ternary & FCC(surf)\tnote{c}, HCP(surf)\tnote{c} & 1000 & 2 & 272.6, 407.3 & $NVT$ & 0.07 \\
94& binary\&ternary & FCC(surf)\tnote{c}, HCP(surf)\tnote{c} & 3000 & 6 & 272.6, 407.3 & $NVT$ & 0.11 \\
95& Al-Cu\&ternary & FCC(surf)\tnote{d}, HCP(surf)\tnote{e} & 1000 & 2 & 50, 135.7 & $NVT$ & 0.06 \\
96& Al-Cu\&ternary & FCC(surf)\tnote{d}, HCP(surf)\tnote{e} & 3000 & 6 & 50, 135.7 & $NVT$ & 0.03 \\
97& Al-Cu\&ternary & FCC(surf)\tnote{d}, HCP(surf)\tnote{e} & 1000 & 2 & 272.6, 407.3 & $NVT$ & 0.23 \\
98& Al-Cu\&ternary & FCC(surf)\tnote{d}, HCP(surf)\tnote{e} & 3000 & 6 & 272.6, 407.3 & $NVT$ & 0.08 \\
\hline
\end{tabular}
\end{threeparttable}}
\end{table}
\end{center}

\begin{center}
\begin{table}[H]
\resizebox{\textwidth}{!}{
\begin{threeparttable}
\begin{tabular}{cccccccc}
\hline
Iter& Alloy & Crystal & DPMD steps & length/ps & T/K & Ensemble & Candidates per/\% \\
\hline
99& Al-Cu\&ternary & FCC(surf)\tnote{d}, HCP(surf)\tnote{e} & 1000 & 2 & 543.1, 678.9 & $NVT$ & 0.21 \\
100& Al-Cu\&ternary & FCC(surf)\tnote{d}, HCP(surf)\tnote{e} & 3000 & 6 & 543.1, 678.9 & $NVT$ & 0.05 \\
101& Al-Cu\&ternary & FCC(surf)\tnote{d}, HCP(surf)\tnote{e} & 1000 & 2 & 814.7, 950.4 & $NVT$ & 0.10 \\
102& Al-Cu\&ternary & FCC(surf)\tnote{d}, HCP(surf)\tnote{e} & 3000 & 6 & 814.7, 950.4 & $NVT$ & 0.03 \\
103& Al-Cu\&ternary & FCC(surf)\tnote{d}, HCP(surf)\tnote{e} & 1000 & 2 & 1086.2, 1222.0 & $NVT$ & 0.12 \\
\hline
\end{tabular}
 \begin{tablenotes}
        \footnotesize
        \item[a] Isothermal-isobaric ensemble with isotropic volume fluctuation.
        \item[b] Isothrtmal-isobaric ensemble with anisotropic cell fluctuation.
        \item[c] Surface conformations generated by displacing conventional cells of bulk crystals.
        \item[d] Surface conformations generated by displacing replicated conventional cells of FCC structures. The $2\times2\times1$ replication is used before displacing the structure along {100} and {110} crystallographic directions. The $2\times1\times1$ replication is used for the {111} direction.
        \item[e] Surface conformations generated by displacing $2\times2\times1$ replicated conventional cells of HCP structures.
      \end{tablenotes}
\end{threeparttable}}
\end{table}
\end{center}
\newpage
\bibliography{ref}
\bibliographystyle{cpb}

%\end{CJK*}  %% end the Chinese environment
\end{document}